\documentstyle[prl,aps]{revtex}
\input BoxedEPS.tex
\SetOzTeXEPSFSpecial
\HideDisplacementBoxes

\begin{document}

\twocolumn[\hsize\textwidth\columnwidth\hsize
           \csname @twocolumnfalse\endcsname
\title{Evidence for the immobile bipolaron formation in
the paramagnetic state of the magnetoresistive manganites}
\author{Guo-meng Zhao$^{(1,2)}$, Y. S. Wang$^{(3)}$, D. J. Kang$^{(1)}$, W.
Prellier$^{(1,*)}$,
M. Rajeswari$^{(1)}$, }
\author{H. Keller $^{(2)}$, T. Venkatesan $^{(1)}$, C. W. Chu$^{(3)}$, and
R. L.
Greene$^{(1)}$ }
\vspace{1cm}
\address{$^{(1)}$Center for Superconductivity Research, University of
Maryland, College Park, MD 20742,
USA\\
$^{(2)}$Physik-Institut der Universit\"at Z\"urich, CH-8057
Z\"urich, Switzerland\\
$^{(3)}$Department of Physics and Texas Center for Superconductivity,
University of Houston, Houston, Texas 77204, USA}

\maketitle
\noindent
\begin{abstract}
Recent research suggests that the charge carriers in
the paramagnetic state of the magnetoresistive
manganites are small polarons.  Here we
report studies of the oxygen-isotope effects on the intrinsic
resistivity and thermoelectric power in several ferromagnetic manganites.
The precise measurements of these isotope effects allow us to make a
quantitative data analysis.  Our results do not support a simple
small-polaron model, but rather
provide compelling evidence for the presence of small immobile bipolarons,
i.e., pairs of small polarons. Since the bipolarons in the manganites
are immobile, the present result alone appears not to give a positive
support to the bipolaronic superconductivity theory for the
copper-based perovskites.
\end{abstract}
\vspace{0.4cm}
]
\narrowtext
The discovery of ``colossal" magnetoresistance (CMR) in thin films of
Re$_{1-x}$A$_{x}$MnO$_{3}$ (Re = a rare-earth element, and A = a divalent
element) \cite{Von} has stimulated extensive studies of magnetic,
structural and
transport properties of these
materials \cite{Art}. The physics of manganites
has primarily been described by the
double-exchange (DE) model \cite{Zener}. However, Millis, Littlewood and
Shraiman \cite{Millis1} pointed out that the carrier-spin
interaction in the DE model is too weak to lead to carrier
localization in the paramagnetic (PM) state, and thus a second mechanism
such as small polaronic effects
should be involved to explain the observed resistivity data
in doped manganites. Following this original idea, more theoretical
models were proposed \cite{Roder,Moreo}. On the other hand, Alexandrov
and Bratkovsky \cite{Alex} have recently shown that the essential physics in
manganites should involve the formation of small bipolarons in the PM state
in order to explain CMR quantitatively.

The first experimental evidence for small polaronic charge carriers in the
PM state was provided by transport measurements \cite{Jaime}. It
was found that the activation energy $E_{\rho}$ deduced from the conductivity
data is one order of magnitude larger than the activation energy $E_{s}$
obtained from the thermoelectric power data. Such a large difference in the
activation energies is the hallmark of the small-polaron hopping
conduction. The giant oxygen-isotope shifts of the ferromagnetic
transition temperature $T_{C}$ give clear evidence for the presence
of polaronic charge carriers in this system \cite{ZhaoNature,ZhaoPRL}.
Moreover, the fast and local techniques have directly shown
that the doped charge carriers are accompanied by local
Jahn-Teller distortions \cite{Billinge,Booth,Louca,Lanzara}.
However, all these experiments cannot make a distinction
between small polarons and small bipolarons since both are dressed by
local lattice distortions. Small bipolarons are
normally much heavier than small polarons, and should be localized in
the presence of small random potentials. In order to discriminate
between polarons and bipolarons and to place constraints on 
the CMR theories, we studied the oxygen-isotope effects on
the intrinsic resistivity in the high-quality
epitaxial thin films of
La$_{0.75}$Ca$_{0.25}$MnO$_{3}$ and
Nd$_{0.7}$Sr$_{0.3}$MnO$_{3}$. We also measured the thermoelectric
power for the oxygen-isotope exchanged ceramic
samples of La$_{0.75}$Ca$_{0.25}$MnO$_{3}$. The data cannot be
explained by a simple small-polaron model, but are in quantitative
agreement with a model where the formation of small immobile bipolarons is
essential.


The epitaxial thin films of La$_{0.75}$Ca$_{0.25}$MnO$_{3}$ (LCMO) and
Nd$_{0.7}$Sr$_{0.3}$MnO$_{3}$ (NSMO) were grown
on $<$100$>$ LaAlO$_{3}$ single crystal substrates by pulsed laser
deposition using a KrF excimer laser \cite{Prellier}.
The film thickness was about 190 nm for NSMO and 150 nm for LCMO.
Two halves were cut
from the same piece of a film for oxygen-isotope diffusion.
The diffusion for LCMO/NSMO was
carried out for 10 h
at about 940/900 $^{\circ}$C and oxygen pressure of 1 bar. The
$^{18}$O-isotope gas is enriched with
95$\%$ $^{18}$O, which can ensure 95$\%$ $^{18}$O in the $^{18}$O thin
films. The ceramic $^{16}$O and $^{18}$O samples of
La$_{0.75}$Ca$_{0.25}$MnO$_{3}$ were the same as those reported in
Ref.~\cite{Zhao99}. The resistivity was measured using the van der Pauw
technique, and the
contacts were made by silver paste. The measurements were
carried out in a Quantum Design measuring system. The thermoelectric
power was measured using an apparatus modeled after a seesaw
technique \cite{Resel}. The absolute uncertainty is less than 0.25 $\mu$V/K
and the
systematic error is $\pm$0.1$\mu$V/K.

Fig.~1 shows the resistivity of the oxygen-isotope exchanged LCMO films
above 1.1$T_{C}$, where $T_{C}$ = 231.5 K for the $^{16}$O sample and
216.5 K for the $^{18}$O sample (see Table 1). The oxygen-isotope shift of
$T_{C}$ is 15.0(6) K in the films, in excellent agreement with
the results for the bulk samples \cite{Zhao99}. From the figure, one
can see that there is a large difference in the intrinsic resistivity
between the two isotope samples. Such a large isotope effect is
reversible upon the oxygen isotope back-exchange. We should mention that
the intrinsic
resistivity of the compounds can be only obtained in high-quality
thin films and single crystals. Our LCMO films even have lower residual
resistivity $\rho_{o}$ than the corresponding single crystals
\cite{ZhaoPRL00},
indicating a high-quality of the films. We also checked that the
resistivity for a ceramic sample of La$_{0.66}$Ba$_{0.34}$MnO$_{3}$ is
very different from that for the corresponding single crystal; the activation
energy $E_{\rho}$ for the former is about 3 times larger than for the
latter. The temperature dependence of the resistivity obtained in the ceramic
samples does not represent the intrinsic behavior of the bulk, and thus
one cannot use ceramic samples to study the isotope effect on the
intrinsic resistivity. Moreover, the van der Pauw technique is
particularly good to precisely measure the resistivity difference between
oxygen-isotope exchanged films whose thicknesses are identical.  Thus
the data shown in Fig.~1 represent the first precise measurements on
the intrinsic resistivity of the two isotope samples.

\begin{figure}[htb]
    \ForceWidth{6.6cm}
	\centerline{\BoxedEPSF{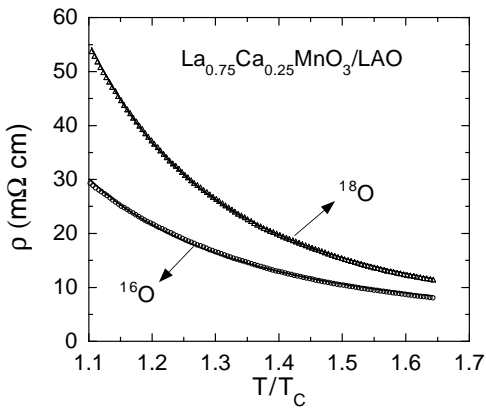}}
	\vspace{0.3cm}
	\caption[~]{The resistivity of the oxygen-isotope exchanged
	films of La$_{0.75}$Ca$_{0.25}$MnO$_{3}$. The maximum temperature of
	the data points for the $^{16}$O film is 380 K. The solid lines are
the
	fitted curves by Eq.~1. As in Ref.~\cite{Jaime}, we
excluded the data points below 1.1$T_{C}$ for the fitting.}
	\protect\label{Fig.1}
\end{figure}

In the PM state, the dominant conduction mechanism is
thermally activated hopping of adiabatic small polarons. When
$T$$>$$W_{p}/k_{B}$ (where $W_{p}$ is the polaron bandwidth), the
temperature dependence of the resistivity is given by \cite{Emin,Austin,Jaime}
\begin{equation}
\rho = BT\exp[(E_{a}+ E_{s})/k_{B}T],
\end{equation}
where $B$ is a coefficient, depending on a
characteristic optical phonon
frequency $\omega_{o}$ (i.e., $B \propto 1/\omega_{o}$); $E_{s}$ is
the energy required to excite
a polaron from a localized impurity state (see Fig.~2a); $E_{a} =
(\eta E_{p}/2) - t$ (here $E_{p}$ is the polaron binding energy, $t$ is
the ``bare'' hopping
integral, $\eta$ = 1 for Holstein polarons
\cite{Emin} and
$\eta$ $\sim$ 0.2-0.4 for Fr\"ohlich polarons \cite{Alexcond}). In the
harmonic approximation, $E_{p}$ and  $E_{a}$ are independent of the
isotope mass $M$. Moreover, since $E_{s}$ does not
depend on $M$ either \cite{Austin}, one expects that $E_{\rho}$ = $E_{a}+
E_{s}$ should be independent of $M$.

The thermoelectric power is given by \cite{Jaime,Austin}
\begin{equation}
S = E_{s}/eT + S_{o}
\end{equation}
where $S_{o}$ is a constant depending on the kinetic energy of the
polarons and on the polaron density \cite{Austin}. One should note
that Eq.~2 is valid only if there is one type of carriers (e.g., holes).
Comparing Eq.~1 and Eq.~2, one readily finds
that $E_{\rho} = E_{a}+ E_{s}$ $>$$>$ $E_{s}$. Jaime {\em et al.}
\cite{Jaime} used the above
equations to fit their resistivity and thermoelectric power data, and
found that $E_{\rho}$$>$$>$ $E_{s}$. This is consistent with the small
polaron hopping mechanism.

We now fit our data by Eq.~1 (see solid lines in Fig.~1). It is
apparent that the fits are quite good for both isotope samples.
However, the isotope dependencies of the parameters $B$ and
$E_{\rho}$ are not expected from Eq.~1. Upon replacing $^{16}$O with $^{18}$O,
the parameter $B$
decreases by 20$\%$, and the parameter $E_{\rho}$ increases by 11
meV. This is in contradiction with Eq.~1 which predicts that the
parameter $B$ should increase by about 6$\%$ while $E_{\rho}$ should not
change. Since our measurements are very accurate, the unusual isotope effect
on $B$ cannot be caused by the experimental uncertainty.
One possible explanation is that Eq.~1 does not hold below
380 K (the maximum temperature of the data points in Fig.~1) since
the condition $T$$>$$W_{p}/k_{B}$ may not be satisfied in
this system.
\begin{figure}[htb]
\ForceWidth{6.6cm}
	\centerline{\BoxedEPSF{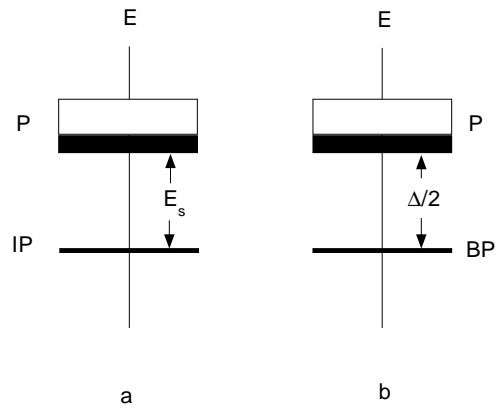}}
	\vspace{0.3cm}
	\caption[~]{A schematic diagram of the polaron band, and polaron
trapping
	into impurity (IP) states (a), or into localized bipolaron (BP) states
	(b).}
	\protect\label{Fig.2}
\end{figure}

One should modify Eq.~1 if $T$$<$$W_{p}/k_{B}$. It is known that the
resistivity can be generally expressed as $\rho = 1/\sigma = 1/ne\mu$,
where $n$ is the mobile carrier concentration and $\mu$ is the
mobility of the carriers. For adiabatic small-polaron hopping, the mobility
is given by \cite{Emin}
\begin{equation}
\mu = \frac{ed^{2}}{h}\frac{\hbar\omega_{o}}{k_{B}T}\exp (-
E_{a}/k_{B}T),
\end{equation}
where $d$ is the site to site hopping distance, which is equal to
$a/\sqrt{2}$ in manganites since the doped holes in this system mainly
reside on the oxygen
sites \cite{Ju}. Here $E_{a}$ should also be modified at low
temperatures \cite{Austin}, i.e., $E_{a} =
(\eta E_{p}/2)f(T) - t$, where $f(T) = [\tanh (\hbar\omega_{o}/4k_{B}T)]/
(\hbar\omega_{o}/4k_{B}T)$ for $T > \hbar\omega_{o}/4k_{B} \simeq$
200 K \cite{Austin}. The mobile polaron density $n$ can be easily
calculated with the help of Fig.~2. When $T$$>$$W_{p}/k_{B}$,
$n \propto \exp(- E_{s}/k_{B}T)$ \cite{Austin}, which leads to Eq.~1
by combining with Eq.~3.
On the other hand, when $T$$<$$W_{p}/k_{B}$,
$n = 2(2\pi m_{p}k_{B}T/h^{2})^{3/2}\exp(- E_{s}/k_{B}T)$ if we
assume a simple parabolic band \cite{Alex,Austin}. Here $m_{p}$ is the
effective mass of polarons and related to $W_{p}$
by $m_{p} = 6\hbar^{2}/a^{2}W_{p}$. In fact,
the above $n(T)$ expression is the same
as that for semiconductors when the chemical potential is pinned to the
impurity
levels. Using the above $n(T)$ expression in the case of $T$$<$$W_{p}/k_{B}$
and Eq.~3,  we finally have
\begin{equation}
\rho = \frac{C}{\sqrt{T}}\exp(E_{\rho}/k_{B}T),
\end{equation}
where $C = (ah/e^{2}\sqrt{k_{B}})(1.05W_{p})^{1.5}/\hbar\omega_{o}$.
The quantity $C$
should strongly depend on the
isotope mass $M$ and de-
\begin{figure}[htb]
\ForceWidth{6.6cm}
	\centerline{\BoxedEPSF{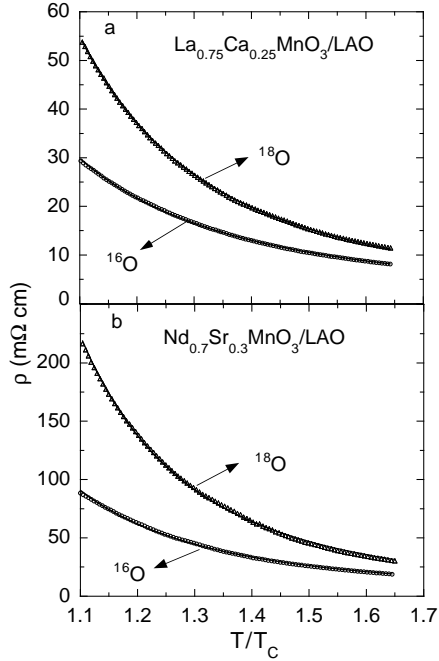}}
	\vspace{0.3cm}
	\caption[~]{The resistivity of the oxygen-isotope exchanged
	films of (a) La$_{0.75}$Ca$_{0.25}$MnO$_{3}$; (b)
Nd$_{0.7}$Sr$_{0.3}$MnO$_{3}$. The solid lines are fitted curves by
Eq.~4.}
	\protect\label{Fig.3}
\end{figure}
\noindent
crease with increasing $M$. This is because  $W_{p}$ decreases
strongly with increasing $M$  according to $W_{p}$ = $W_{o}\exp(-\Gamma
E_{p}/\hbar\omega_{o})$ = $
W_{o}\exp(-g^{2})$ \cite{Alex99,Alexcond}. Here $W_{o}$ is the bare
bandwidth ($W_{o}$ = 12$t$), $\Gamma$ is
an isotope independent constant, which is less than 0.4 for both Holstein
and Fr\"ohlich
polarons when the coupling constant $\lambda = 2E_{p}/W_{o}$$<$0.5
and $\hbar\omega_{o}/t$$<$1 \cite{Alex99}. Clearly, the parameter
$C$ should decrease significantly with increasing oxygen-isotope mass,
in contrast to the parameter $B$ ($\propto
1/\omega_{o}$) which would increase by about 6$\%$ upon replacing $^{16}$O
with $^{18}$O.

Now if one considers that
small polarons are bound into localized bipolarons (i.e., pairs of
small polarons) \cite{Alex}, the only
modification for both Eq.~2 and Eq.~4 is to replace $E_{s}$ by $\Delta$/2
(see Fig.~2). Here $\Delta$ is
the bipolaron binding energy and given by $\Delta = 2(1-\Gamma)E_{p}- V_{c}-
W_{p}$, where $V_{c}$ is the Coulombic repulsion between
bound polarons \cite{Alexcond}. It is apparent that $\Delta$ and
thus $E_{s}$ should depend
on the isotope mass $M$ due to the fact that $W_{p}$ is a strong
function of $M$.

\begin{figure}[htb]
    \ForceWidth{6.6cm}
	\centerline{\BoxedEPSF{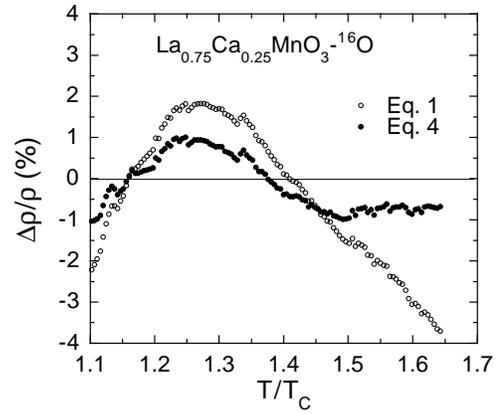}}
	\vspace{0.3cm}
	\caption[~]{The deviations between the data of the $^{16}$O film of
	La$_{0.75}$Ca$_{0.25}$MnO$_{3}$ and the fitted curves by
	Eq.~1 and Eq.~4.}
	\protect\label{Fig.4}
\end{figure}

In Fig.~3, we show the resistivity data for the
$^{16}$O and $^{18}$O films of both LCMO and NSMO compounds, and fit
them by Eq.~4. The data can be
well fitted by Eq.~4 with the parameters summarized in Table I. It is
striking that the fits with Eq.~4 are much better than the fits
with Eq.~1, as seen more clearly in Fig.~4 where the deviations
between the data and the fitted curves are plotted for the LCMO $^{16}$O
film. The fit with Eq.~1 has a large systematic deviation ($>$4$\%$),
while the fit with Eq.~4 has a much smaller deviation ($<$1$\%$).
It is easy to check that the $T$
dependence of the prefactor of the
exponential function in Eq.~1 and Eq.~4 is not important when
$E_{\rho}$ is large, but becomes crucial when $E_{\rho}$ is small.
This can naturally explain why the resistivity data in
the La$_{1-x}$Ca$_{x}$MnO$_{3}$ system can be well fitted by Eq.~1, but the
data of the La$_{0.7}$Sr$_{0.3}$MnO$_{3}$ film are not consistent with
Eq.~1  \cite{Snyder}. We speculate that the
data of La$_{0.7}$Sr$_{0.3}$MnO$_{3}$ in Ref.~\cite{Snyder} should
be in better agreement with Eq.~4.

From Table I, one can see that the parameter $C$ for LCMO decreases by
35(5)$\%$, and $E_{\rho}$ increases by 13.2(3) meV. For NSMO, $C$ decreases by
40(7)$\%$, and $E_{\rho}$ increases by 14.2(8) meV. The deduced oxygen-isotope
effects on the parameter $C$ and $E_{\rho}$ are in qualitative
agreement with Eq.~4. Since this equation is valid independent of whether
the small polarons are bound into
localized bipolarons or to the impurity centers (see Fig.~2 and
discussion above), one can only draw a
conclusion that the isotope dependence of the electrical transport in
the PM state of the
manganites agrees qualitatively with small-polaron hopping conduction.

Now the question arises: Can the present data tell whether
the small polarons are bound into
localized bipolarons or to the impurity centers?  The clarification
of this issue can place an essential constraint on various CMR
theories and on bipolaronic superconductivity theory for the
cuprate superconductors. If the small polarons are bound to impurity centers,
there will be no isotope effect on $E_{s}$ \cite{Austin}. Then, the isotope
effect on $E_{\rho}$ only arises from the isotope shift of $E_{a}$
because $E_{\rho} = E_{a} + E_{s}$. Since
$E_{a} = (\eta E_{p}/2)f(T) - W_{o}/12$, only the quantity
$f(T) = [\tanh (\hbar\omega_{o}/4k_{B}T)]/
(\hbar\omega_{o}/4k_{B}T)$ may depend on the isotope mass if the
temperature is not so high compared with $\hbar\omega_{o}/k_{B}$. If we
take $\hbar\omega_{o}$ = 74 meV, typical for oxides \cite{Alex99},
we have $f(T)$ = 0.855 and the isotope-induced change $\delta f(T)$ = 0.0134
at T = 300 K, the midpoint temperature of the data points for the LCMO
films. Using
the relation: $(\eta E_{p}/2)f(T) = W_{o}/12 + E_{\rho} - E_{s}$, and
$E_{s}$ = 13.2 meV for
La$_{0.75}$Ca$_{0.25}$MnO$_{3}$ (see below), we find $\eta
E_{p}/2$ = 0.55 eV. Here we have used $W_{o}$ = 4.9 eV, as estimated from
the relation: $W_{o} = 6\hbar^{2}/a^{2}m_{b}$
and $m_{b}$ = 0.61$m_{e}$ \cite{Zhaopreprint}. The estimated bare
bandwidth is typical for the oxygen band and in good agreement with
the electron-energy-loss spectra \cite{Ju}. Then $\delta E_{\rho} = \delta
E_{a}
= (\eta E_{p}/2)\delta f(T)$ = 7.4 meV, which is about half the value
observed. This implies that there must be an isotope effect on
$E_{s}$, which is only possible if the small polarons are bound into
localized bipolarons as discussed above. The oxygen-isotope shift of
$E_{s}$ in the LCMO compound is
$\delta E_{s}$ = $\delta E_{\rho}$ - $\delta E_{a}$ = 5.8(3)
meV.
\begin{table}[htb]
	\caption[~]{Summary of the fitting and measured parameters for the
	$^{16}$O and $^{18}$O films of La$_{0.75}$Ca$_{0.25}$MnO$_{3}$ (LCMO)
	and Nd$_{0.7}$Sr$_{0.3}$MnO$_{3}$ (NSMO). The errors of the
	parameters comes from the fitting and from the van der Pauw
measurement.
	The absolute uncertainty of the thickness of the films was not
	included in the error calculations since it only influences the
	absolute values of the resistivity.}
	\begin{center}
    \begin{tabular}{lccccc}
    	Compounds & $T_{C}$&
    	$\rho_{0}$ & $C$ &$E_{\rho}$
    	 \\
    	 &(K) &($\mu\Omega$cm)&(m$\Omega$cmK$^{0.5}$) &
    	 (meV)\\
    	\hline
    	LCMO($^{16}$O)&231.5(3) &122(2)&17.3(5) & 72.8(2)\\
    	LCMO($^{18}$O)&216.5(3)&141(2)& 12.9(3)& 86.0(1))\\
    NSMO($^{16}$O)&204(1)& 248(4)&23.2(8) & 78.8(4)\\
    NSMO($^{18}$O)&186(1)&289(4)&16.2(7) & 92.9(4)\\
\end{tabular}
\end{center}
\protect\label{Tab1}
\end{table}
One can also obtain the isotope shift of $E_{s}$ by
measuring the thermoelectric power for two isotope samples according
to Eq.~2.  In Fig.~5, we plot the thermoelectric power $S$ as a
function of $1/T$ for the $^{16}$O and $^{18}$O samples of
La$_{0.75}$Ca$_{0.25}$MnO$_{3}$. Both $T_{C}$'s and the isotope shift
of the ceramic samples \cite{Zhao99} are the same as those in the
corresponding
thin films. Since the grain-boundary effect on
$S$ is negligible, the thermoelectric power obtained in ceramic samples
should be intrinsic. From the slops of the straight lines in
Fig.~5, we find $E_{s}$ = 13.2 meV for the $^{16}$O sample and 18.7
meV for the $^{18}$O. The isotope shift is $\delta E_{s}$ = 5.5(5) meV,
in remarkably good agreement with that (5.8 meV) deduced independently from
the
above resistivity data.

\begin{figure}[htb]
\ForceWidth{6.6cm}
	\centerline{\BoxedEPSF{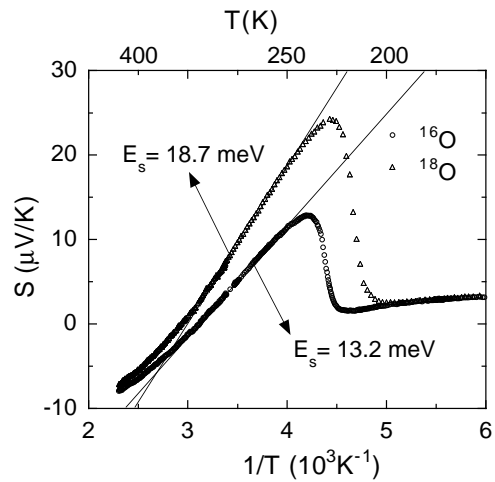}}
	\vspace{0.3cm}
	\caption[~]{The thermoelectric power $S(T)$ of the oxygen-isotope
exchanged
	ceramic samples of La$_{0.75}$Ca$_{0.25}$MnO$_{3}$.}
	\protect\label{Fig.5}
\end{figure}

We can use the values of the parameter $C$ for the $^{16}$O films to calculate
the polaron bandwidth $W_{p}$ according to the
relation: $C = (ah/e^{2}\sqrt{k_{B}})(1.05W_{p})^{1.5}/\hbar\omega_{o}$.
By taking $\hbar\omega_{o}$ = 74 meV, we obtain $W_{p}$ = 49(2) meV for
the LCMO $^{16}$O film, and 60(3) meV for the NSMO $^{16}$O film. A
larger $W_{p}$ for the NSMO film might be an artifact since the
residual resistivity $\rho_{o}$ of this film is about 40$\%$ larger
than that of the best single crystal \cite{Sawaki}. The discrepancy
is possibly due to the fact that the interdiffusion between the NSMO
film and substrate might occur during the high-temperature anneal.
From the $W_{p}$ values, one can see that for our data $T < W_{p}/k_{B}$,
which
justifies the use of
Eq.~4. Moreover, the polaron
bandwidth is greatly reduced compared with the bare bandwidth
$W_{o}$ = 4.9 eV. Using $g^{2} = \ln (W_{o}/W_{p})$, we can
determine $g^{2}$ to be 4.6(1)/4.4(3) for the LCMO/NSMO $^{16}$O film.
If $\hbar\omega_{o}$ decreases by 5.7$\%$
upon replacing $^{16}$O by
$^{18}$O, then our calculation shows that the parameter $C$ decreases
by 35$\%$ for LCMO, and by 33$\%$ for the
NSMO, in good agreement with the measured values:
35(5)$\%$ for LCMO and 40(7)$\%$ for NSMO.

Furthermore, one can quantitatively explain the
isotope dependence of $E_{\rho}$ if small polarons form localized
bipolarons. In this scenario, $\delta\Delta = - \delta W_{p}$ (see
above), so $\delta\Delta  = 0.057g^{2}W_{p}$.
From the deduced values for $g^{2}$ and $W_{p}$ above, we
calculate that $\delta\Delta$ = 12.9 meV for LCMO, and 15.0 meV for
NSMO. So $\delta E_{s}$ = $\delta\Delta$/2 = 6.5 meV for LCMO, in quantitative
agreement with the values deduced independently from both resistivity and
thermoelectric power data. Using $\delta E_{\rho} =
\delta E_{a} + \delta\Delta /2$, we find $\delta E_{\rho}$= 13.8
meV for LCMO, and 15.3 meV for NSMO. The calculated values are in excellent
agreement with the observed values: 13.2(3) meV for LCMO and 14.2(8) meV
for NSMO.

Since the bipolarons in the manganites are immobile, it is natural to
ask whether bipolarons in the cuprate superconductors, if exist, would
be mobile and responsible for high-temperature superconductivity
\cite{Alexbook}. The bipolarons in cuprates would be mobile if the
electron-phonon coupling in this system were much weaker than in
manganites. This appears not to be the case since the magnitudes of
the long-range
Fr\"ohlich electron-phonon interaction in both systems are similar
\cite{Alexcond}, and the short-range Jahn-Teller (JT) electron-phonon
coupling in cuprates is even much stronger than that in manganites
(in La$_{2}$CuO$_{4}$, the Jahn-Teller stabilization
energy $E_{JT} \simeq$ 1.2 eV for the $Q_{3}$-type mode \cite{Kamimura}, while
$E_{JT} \simeq$ 0.5 eV
in LaMnO$_{3}$ \cite{Millis1}). The bipolarons can become mobile only
if the $Q_{3}$-type ``anti-JT'' mode in cuprates is not active and
does not lead to the formation of anti-JT polarons possibly due to the too
large
$E_{JT}$. Experimentally, the
dynamic JT distortions (or anti-JT polarons) have
been observed in doped manganites
\cite{Billinge,Booth,Louca,Lanzara}, while the dynamic $Q_{3}$-type
JT distortions have so far not been observed in doped cuprates.
Therefore, our present result alone appears not to give a positive support
to the
bipolaronic superconductivity in cuprates unless both experiment and
theory can clearly show that anti-JT polarons ($Q_{3}$-type) cannot be
formed in this system.

In summary, we have observed large oxygen isotope effects on the intrinsic
resistivity in high-quality
epitaxial thin films of
La$_{0.75}$Ca$_{0.25}$MnO$_{3}$ and Nd$_{0.7}$Sr$_{0.3}$MnO$_{3}$, and
on the thermoelectric power in the ceramic samples of
La$_{0.75}$Ca$_{0.25}$MnO$_{3}$. The data can
be quantitatively explained by a scenario \cite{Alex} where the small polarons
form localized bound pairs (bipolarons) in the paramagnetic state. The
coexistence of small polarons and bipolarons in the PM state may lead
to a dynamic phase separation into the insulating
antiferromagnetically coupled region where the bipolarons reside, and
into the ferromagnetically coupled region where the polarons
sit. This simple picture can naturally explain the observation of the
ferromagnetic clusters in the PM state \cite{Teresa}. Although we
cannot completely rule out the possibility that the other models might
also be able to explain the present isotope effects, we would like to
point out that the agreement between theory and experiment should be
quantitative.

{\bf Acknowlegment}: We would like to thank A. S. Alexandrov and A. M.
Bratkovsky for useful discussions. We also thank X. M. Zhang for
helping with resistivity measurements. The work was supported by the
NSF MRSEC at the University of Maryland and Swiss National Science
Foundation.
~\\
~\\
$*$ Present address: Laboratoire CRISMAT-ISMRA,14050 CAEN Cedex, France.
\bibliographystyle{prsty}



\end{document}